\documentstyle[twocolumn,epsfig]{mn}
\begin{document}

\title{The role of Kelvin-Helmholtz instability in  dusty and partially ionized outflows}

\author[M. Shadmehri and T. P. Downes]{Mohsen Shadmehri$^{1,2}$\thanks{E-mail:
mohsen.shadmehri@dcu.ie (MS); } and Turlough P. Downes$^{1}$\thanks{E-mail:
turlough.downes@dcu.ie (TPD)}\\
$^{1}$School of Mathematical Sciences, Dublin City University, Glasnevin, Dublin 9, Ireland\\
$^{2}$ Department of Physics, School of Science, Ferdowsi University, Mashhad, Iran}

\maketitle

\date{Received ______________ / Accepted _________________ }

\begin{abstract}
We investigate the linear theory of Kelvin-Helmholtz instability at the interface between  a partially ionized dusty outflow and the ambient material  analytically. We model the interaction as a multifluid system in a planar geometry. The unstable modes are independent from the charge polarity of the dust particles. Although our results show a stabilizing effect for charged dust particles, the growth time scale of the growing modes gradually becomes independent of the mass or charge of the dust particles when the magnetic field strength increases. We show that growth time scale decreases with increasing the magnetic field. Also, as the mass of the dust particles increases, the growth time scale of the unstable mode increases.
\end{abstract}

\section{Introduction}
\label{sec:1}

The Kelvin-Helmholtz (KH) instability is of much interest in the investigation of variety of
the astrophysical phenomena such as interaction between jets or outflows and the
ambient medium (e.g., Watson et al. 2007; Birk \& Wiechen 2002; Rosen et al. 1999; Downes \& Ray 1998; Hardee \& Stone 1997; Bodo et al. 1995; ). This
instability, the simplest example of shear flow instability, is a
well-known phenomenon in fluid mechanics and astrophysics (e.g.\
Chandrasekhar 1961). Despite the very existence of the ions, neutrals and charged dust particles in jets or outflows (Markwick-Kemper, Green \& Peeters 2005; Weinberger \& Armsdorfer 2004; Gueth, Bachiller \& Tafalla 2003; Shepherd 2001), the KH instability in these systems has been studied mostly in one fluid approximation just for simplicity. Dusty outflows are observed in some of the starburst galaxies (e.g., Alton, Davies \& Bianchi 1999). However, a two-fluid treatment of KH instability, taking
account of the different motions of ions and neutrals and of the
magnetic field, has been investigated by some authors as well (e.g., Birk et al. 2000; Watson et al. 2004; Chhajlani \& Vyas 1991; Chhajlani \& Vyas 1990).

Birk et al.\ (2000) studied the KH
instability by taking into account the full incompressible dynamics of
both the neutral and the ionized gas components with applications to
multi-phase galactic outflow winds. In another similar study, Watson et
al.\ (2004)  investigated the KH instability in the
linear, partially ionized regime to determine its possible effect on entrainment in massive bipolar outflows. They showed that for much of the relevant parameter space, neutral and ions are sufficiently decoupled that the neutrals are unstable while ions are held in place by the magnetic field. Shadmehri \& Downes (2007) extended this analysis to a {\it layer} of ions and neutrals with finite thickness. They showed that perturbations with wavelength comparable to layer's thickness are significantly affected by the
thickness of the layer. Birk \& Wiechen (2002) focused on unstable shear flows in partially ionized dense dusty plasma. They considered the dust and neutral gas components so that dust and neutral collisions is the dominant momentum transfer mechanism and dust component can interact with magnetic field lines, although dust charge fluctuations are negligible. They showed long wavelength modes can be stabilized by dust and neutral gas collisional momentum transfer. However, their analysis focused on numerical solution and multifluid numerical simulations of the problem of shear flow in weakly ionized, magnetized dusty plasmas, and did not concentrate on solving the dispersion relation for this type of flow. In this paper, we will obtain the dispersion relation of KH instability in a multifluid system.

Recently, Wiechen (2006) studied the KH instability by doing   multifluid numerical simulations in partially ionized, dusty plasmas for different masses and charges of dust. He showed a stabilizing effect for more massive dust grains with no dependence on the charge polarity of the dust. In this paper, we study KH instability in a multifluid system {\it analytically}. In particular, we apply our results to magnetized dusty outflows. In the next section, the basic equations and assumptions which are based on Pandey \& Vladimirov (2007) are presented. We develop analytic estimates of the relevant time scales of the KH modes. In section 3, properties of the unstable growing modes are studied.

\section{General Formulation}
\label{sec:2}

\subsection{Basic multifluid equations}

Our basic equations and the main assumptions are similar to Pandey \& Vladimirov (2007).
We take account of the different bulk velocities and densities of the neutral, electrons, ions and charged dust particles on both sides of the KH interface. The continuity equation is
\begin{equation}
\frac{\partial\rho_{j}}{\partial t}+\nabla.(\rho_{j} {\bf v}_{j})=0,\label{eq:conj}
\end{equation}
where $\rho_{j}$ and ${\bf v}_{j}$ are the velocity of the various plasma components and the neutrals, respectively.

The momentum equations are
\begin{equation}
0=-q_{j}n_{j}({\bf E'}+ \frac{{\bf v}_{j}\times {\bf B}}{c})-\rho_{j}\nu_{jn}{\bf v}_{j},\label{eq:momj}
\end{equation}
\begin{equation}
\rho_{n}(\frac{\partial {\bf v}_{n}}{\partial t}+ {\bf v}_{n}.\nabla {\bf v}_{n})=-\nabla P + \sum_{e,i,d} \rho_{j}\nu_{jn}{\bf v}_{j}.\label{eq:momn}
\end{equation}
Note that velocities ${\bf v}_{j}$ are written in the neutral frame and ${\bf E'}={\bf E}+ {\bf v}_{n} \times {\bf B}/c$ is the electric field in the  neutral frame. $j$ stands for electrons ($q_{e}=-e$), ions ($q_{i}=e$) and dust ($q_{d}=Ze$), where $Z$ is the number of charge on the grain. The other physical variables have their usual meanings.

Also, the collision frequencies is (Draine et al. 1983)
\begin{equation}
\nu_{jn}=\frac{<\sigma v>_{jn}}{m_{j}+m_{n}}\rho_{n},\label{eq:nu}
\end{equation}
where $<\sigma v>_{jn}$ is the rate coefficient for the momentum transfer by the collision of the $j^{\rm th}$ particle with the neutrals:
\begin{equation}
<\sigma v>_{in}=1.9\times 10^{-9} {\rm cm^{3} s^{-1}},
\end{equation}
\begin{equation}
<\sigma v>_{en}=4.5\times 10^{-9} (\frac{T}{30 \rm K})^{\frac{1}{2}} {\rm cm^{3} s^{-1}},
\end{equation}
and for small grains, we have $<\sigma v>_{dn} \approx <\sigma v>_{in}$, but for grains ranging between a few Angstrom to a few microns (Nakano \& Umebayashi 1986)
\begin{displaymath}
<\sigma v>_{dn}=2.8\times 10^{-5} (\frac{T}{30 \rm K})^{\frac{1}{2}}
\end{displaymath}
\begin{equation}
\times (\frac{a}{10^{-5} \rm cm})^{2} {\rm cm^{3} s^{-1}},\label{eq:dn}
\end{equation}
where $a$ is the grain radius. In our calculation, for the ion mass and mean neutral mass we adopt $m_{i}=30 m_{p}$ and $m_{n}=2.33 m_{p}$, where $m_{p}=1.67 \times 10^{-24} \rm g$ is the proton mass.

Defining the mass density of the bulk fluid and the bulk velocity as $\rho \approx \rho_{n}$ and ${\bf u} \approx {\bf v}_{n}$, equations (\ref{eq:conj}), (\ref{eq:momj}) and (\ref{eq:momn}) give the continuity and the momentum equations for the bulk fluid as
\begin{equation}
\frac{\partial \rho}{\partial t} + \nabla . (\rho {\bf u}) =0,\label{eq:main1}
\end{equation}
\begin{equation}
\rho (\frac{\partial {\bf u}}{\partial t}+ {\bf u}. \nabla {\bf u})=
- \nabla P + \frac{{\bf J}\times {\bf B}}{c}.\label{eq:main2}
\end{equation}

The next simplifying assumption is that electrons and ions are assumed well coupled to the magnetic field which implies $\beta_{e} \gg \beta_{i} \gg 1$, where $\beta_{j}=\omega_{cj}/\nu_{jn}$ is the ratio of cyclotron $\omega_{cj}=q_{j}B/m_{j}c$ to the collision frequencies. Based on this assumption and using quasi-neutrality condition, we have (Pandey \& Vladimirov 2007; Ciolek \& Mouschovias 1993)
\begin{equation}
{\bf v}_{e}=-\frac{1+\Theta}{Zen_{d}}{\bf J},
\end{equation}
where
\begin{equation}
\Theta=[1+\frac{\nu_{nd}}{\nu_{ni}}]\beta_{d}^{2}.\label{eq:Theta}
\end{equation}
Thus, the induction equation can be written as (Pandey \& Vladimirov 2007)
\begin{equation}
\frac{\partial {\bf B}}{\partial t} = \nabla \times [({\bf u}\times{\bf B}) -\frac{1+\Theta}{Zen_{d}} {\bf J}\times {\bf B} ]\label{eq:induc}
\end{equation}

Equations (\ref{eq:main1}), (\ref{eq:main2}) and  (\ref{eq:induc}) along with the equation
\begin{equation}
\nabla . {\bf B}=0
\end{equation}
are our basic equations for Kelvin-Helmholtz instability in a partially ionized medium.  However, we assume incompressibility for the analytical calculations.

\subsection{Linear perturbations}

For doing linear analysis, the unperturbed properties of the system are important. We suppose that the streaming takes place in the $x-$direction with velocity $U(z)$,
\[U(z) = \left\{
\begin{array}{l l}
  +U & \quad \mbox{for $z<0$}\\
  -U & \quad \mbox{for $z>0$}\\ \end{array} \right. \]
where $U$ is constant. The magnetic field is assumed parallel to the
interface; that is, ${\bf B} = B {\bf e}_{x}$. Finally, all
unperturbed physical quantities are assumed constant in each medium.

Now, we can linearize the basic equations. We perturb the physical variables as
\begin{equation}
\chi (z,x,t) = \chi'(z) \exp[i(\omega t + k_{x}x+k_{y}y)].
\end{equation}
Thus,
\begin{equation}
ik_{x}u_{x}' + i k_{y} u_{y}' + \frac{du_{z}'}{dz}=0,\label{eq:con}
\end{equation}
\begin{equation}
\phi \rho u_{x}'=-k_{x}P',\label{eq:xmom}
\end{equation}
\begin{equation}
\phi \rho u_{y}'= -k_{y} P' + \frac{B}{4\pi} (k_{x}
B_{y}'-k_{y}B_{x}'),\label{eq:ymom}
\end{equation}
\begin{equation}
\phi \rho u_{z}'=i\frac{dP'}{dz}-\frac{i B}{4\pi}(ik_{x}
B_{z}'-\frac{dB_{x}'}{dz}),\label{eq:zmom}
\end{equation}
\begin{equation}
\phi B_{x}'=k_{x} B u_{x}' + \eta k_{x} B (\frac{dB_{y}'}{dz}-ik_{y}
B_{z}'),\label{eq:Bx}
\end{equation}
\begin{equation}
\phi B_{y}'= k_{x}B u_{y}'-\eta k_{x}B
(\frac{dB_{x}'}{dz}-ik_{x}B_{z}'),\label{eq:By}
\end{equation}
\begin{equation}
\phi B_{z}'= k_{x} B u_{z}' - \eta k_{x}B i (k_{x}B_{y}'-k_{y}
B_{x}'),\label{eq:Bz}
\end{equation}
\begin{equation}
ik_{x} B_{x}' + ik_{y} B_{y}' + \frac{dB_{z}'}{dz} =0,
\end{equation}
where
\begin{equation}
\phi = \omega + k_{x}U, \eta=\frac{c}{4\pi}\frac{1+\Theta}{Zen_{d}}=\frac{1}{4\pi}\frac{B}{\rho \omega_{mcd}},
\end{equation}
and $\omega_{mcd}=(\rho_{d}/\rho)(1/1+\Theta)(ZeB/m_{d}c)$ is the modified dust-cyclotron frequency.

Now, we can simplify the above  differential equations. By
multiplying equation (\ref{eq:xmom}) by $k_{x}$ and equation
(\ref{eq:ymom}) by $k_{y}$ and then adding the resulting equations,
we obtain
\begin{equation}
i\phi\rho\frac{du_{z}'}{dz}=-k^2 P' + \frac{k_{y}B}{4\pi}(k_{x}
B_{y}'- k_{y}B_{x}'),
\end{equation}
where $k^2=k_{x}^{2}+k_{y}^{2}$. This equation gives $P'$ in terms
of the other variables and the parameters. By substituting $P'$ into
equation (\ref{eq:zmom}), we obtain
\begin{equation}
\rho\phi ({\mathcal{D}}-k^2) u_{z}'=\frac{k_{x}B}{4\pi}
({\mathcal{D}}-k^{2}) B_{z}',\label{eq:MAIN1}
\end{equation}
where ${\mathcal{D}}\equiv \frac{d^{2}}{dz^{2}}$.

On the other hand, using equations (\ref{eq:xmom}) and
(\ref{eq:ymom}) we have
\begin{equation}
k_{x}u_{y}'-k_{y}u_{x}'= \frac{k_{x}B}{4\pi \rho \phi} (k_{x}
B_{y}'- k_{y} B_{x}').\label{eq:help2}
\end{equation}
Also, using equation (\ref{eq:Bx}) and (\ref{eq:By}) we obtain
\begin{displaymath}
k_{x}B_{y}'-k_{y}B_{x}'=\frac{k_{x}B}{\phi}
(k_{x}u_{y}'-k_{y}u_{x}')
\end{displaymath}
\begin{equation}
-\frac{\eta
k_{x}Bi}{\phi}({\mathcal{D}}-k^2)B_{z}'.\label{eq:help3}
\end{equation}
So, from equations (\ref{eq:help2}) and (\ref{eq:help3}) we have
\begin{displaymath}
k_{x}B_{y}'-k_{y}B_{x}' = -i \frac{(\eta k_{x} B /\phi)}{1-
(k_{x}^{2}B^2/4\pi \rho \phi^{2})}
\end{displaymath}
\begin{equation}
\times ({\mathcal{D}} -k^{2})B_{z}'.
\end{equation}
Substituting this equation into equation (\ref{eq:Bz}) gives
\begin{equation}
B_{z}'=\frac{k_{x}B}{\phi} u_{z}'-\frac{(\eta^{2} k_{x}^{2} B^{2}
/\phi^{2})}{1- (k_{x}^{2}B^2/4\pi \rho \phi^{2})}
({\mathcal{D}}-k^{2}) B_{z}'.\label{eq:MAIN2}
\end{equation}
Equations (\ref{eq:MAIN1}) and (\ref{eq:MAIN2}) are our main
equation for KH analysis. However, we can simply them into one
differential equation for $u_{z}'$,
\begin{equation}
({\mathcal{D}}-q^2)({\mathcal{D}}-k^2) u_{z}'=0,\label{eq:final}
\end{equation}
where
\begin{equation}
q^{2}=k^{2}-\frac{1}{4\pi\rho\eta^{2}}(\frac{v_{A}k_{x}}{\phi})^{2}[(\frac{\phi}{v_{A}k_{x}})^{2}-1]^{2},\label{eq:q}
\end{equation}
and $v_{A}=B/\sqrt{4\pi\rho}$ is Alfven  speed.

\subsection{Dispersion relation}
The general solution of equation (\ref{eq:final}) is a linear
combination of $\exp(\pm kz)$ and $\exp(\pm q z)$. Thus, we can write the solutions of equation (\ref{eq:final}) appropriate to the two regions as
\[u_{z}'(z) = \left\{
\begin{array}{l l}
  A_{1} e^{+kz} + A_{2} e^{+q_1 z} & \quad \mbox{for $z<0$}\\
  A_{3} e^{-kz} + A_{4} e^{-q_{2} z} & \quad \mbox{for $z>0$}\\ \end{array} \right. \]
where $A_1$, $A_2$, $A_3$ and $A_4$ are the constants of integration yet to be obtained from
the boundary conditions appropriate to our system. Note that $q_1$ and $q_2$ are given by equation (\ref{eq:q}), and are assumed to have a positive real part so as to render the perturbations bounded at infinity. We have
\begin{equation}
q_{1}^{2} = k^{2}\{1 - (\frac{\omega_{mcd}}{kU})^2
[\frac{{\mathcal M}^2 (x+1)^{2}-1}{x+1}]^{2}\},
\end{equation}
\begin{equation}
q_{2}^{2} = k^{2}\{ 1 - (\frac{\omega_{mcd}}{kU})^2
[\frac{{\mathcal M}^2 (x-1)^{2}-1}{x-1}]^{2}\},
\end{equation}
and ${\mathcal M}= U/ v_{A}$ and $x=\omega/k_{x}U$.

The boundary conditions to be satisfied at the interface $z=0$ are the standard conditions which have been used by many authors (Hunter \& Whitaker  1989; Roychoudhury \& Lovelace 1986; Mehta \& Bhatia 1988; Chhajlani \& Vyas 1990; Chhajlani \& Vyas 1991). We are using the continuity of normal and tangential components of the magnetic field, the vertical displacement of the interface and the total pressure as the boundary conditions. After doing mathematical manipulations, we obtain

\begin{equation}
\frac{A_{1}+A_{2}}{x+1}=\frac{A_{3}+A_{4}}{x-1},
\end{equation}
\begin{equation}
[1- {\mathcal M}^2 (x+1)^{2}]A_{2}=\frac{x+1}{x-1}[1- {\mathcal M}^2
(x-1)^{2}]A_{4},
\end{equation}
\begin{equation}
\frac{x+1}{x-1} (kA_{1} + q_{1} A_{2}) = -kA_{3} - q_{2} A_{4},
\end{equation}
\begin{displaymath}
k A_{1}+q_{1} {\mathcal M}^{2} (x+1)^{2} A_{2}=\frac{x+1}{x-1}
\end{displaymath}
\begin{equation}
\times [-kA_3 - q_{2} {\mathcal M}^{2} (x-1)^{2}A_4].
\end{equation}
For a non-trivial solution of the above algebraic equations, the determinant of the coefficients should vanish. We get the following equation
\begin{equation}
\beta\alpha^{2} q_{1} + q_{2} = \frac{(1-\beta)(\alpha^2 -1 )}{(1+\beta)[1-M^{2}(x+1)^{2}]}k,\label{eq:diss}
\end{equation}
where $\alpha=(x+1)/(x-1)$ and $\beta = [1-M^{2}(x-1)^{2}]/[1-M^{2}(x+1)^{2}]$.  Equation (\ref{eq:diss}) is the general dispersion relation. However, it would be extremely unwieldy to solve it its present form. After substituting the values of $q_1$ and $q_2$ and doing long algebraic mathematical manipulations, we can transform equation (\ref{eq:diss}) into a polynomial form, i.e.

\begin{displaymath}
P_8z^8+P_7z^7+P_6 z^6+P_5 z^5+P_4 z^4+P_3 z^{3}+P_2 z^2
\end{displaymath}
\begin{equation}
+P_1 z + P_0 = 0,\label{eq:dissal}
\end{equation}
where $x=\pm \sqrt{z}/M$ and the coefficients $P_1$ ... $P_8$ are presented in the Appendix. We note that not all  roots of the equation (\ref{eq:dissal}) are acceptable. All roots are inserted into our fundamental dispersion relation (\ref{eq:diss}) to determine which are the valid solutions. We accept only those solutions which have negative imaginary part {\it and} give us positive values for the real parts of $q_1$ and $q_2$.

\begin{figure}
\epsfig{figure=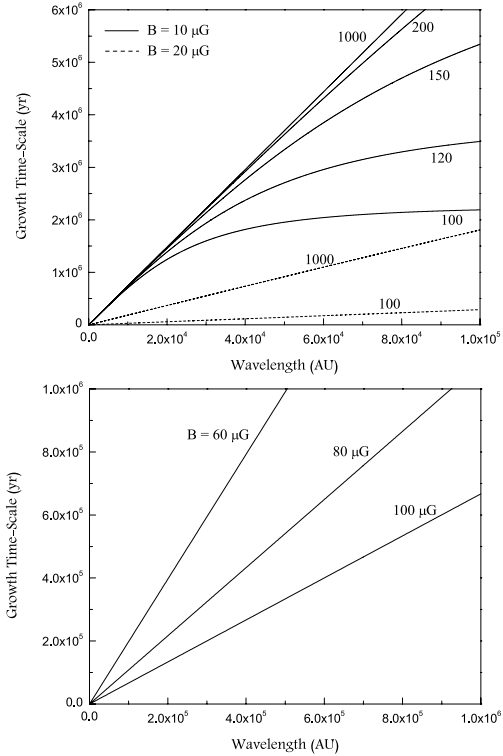,angle=0,width=8.0cm}
\caption{Growth time scales of the instability vs. wavelength of the perturbations for different level of the magnetic strength and various sizes of dust particles. In these plots, it is assumed $|Z|=1$. Curves of top plot are labeled by the size of the dust particles in Angstrom. But since for $B>50 \mu$G, the curves are not very sensitive to the size of the dust particles or their electrical charge, curves of the bottom plot are labeled by the magnetic strength only.}
\label{fig:f1}
\end{figure}

\section{Analysis}
Now, we carry out a parameter study of the roots of equation (\ref{eq:dissal}) as functions of the input parameters. Although this dispersion relation is applicable to a wide range of astrophysical systems, we restrict our analysis to dusty outflows with small grains in order to illustrate the effect of charged dust particles on the growth rate of the unstable modes. We take the molecular hydrogen density to be $10^3$ cm$^3$ and the ratio of ionized to neutral mass density is $10^{-6}$. Also, the ratio of dust to neutral density is assumed to be 0.01. The bulk  density of a dust particle is assumed $1.25$ gr cm$^{-3}$. Also, the relative velocity $2U$ is assumed to be 20 Km s$^{-1}$. We note that the electrical charge of the dust particles  appears only through the modified dust-cyclotron frequency $\omega_{mcd}$. But in our final dispersion relation, this parameter $f=(\omega_{mcd}/kU)^{-2}$ appears and so, the polarity of the electrical charge of the dust particle is not important.

Since we fix the densities and the   velocity for the flow, when the magnetic field strength increases the Mach number decreases. Another important input parameter is modified dust-cyclotron frequency $\omega_{mcd}$ which is calculated based on the other input parameters. Figure \ref{fig:f1} shows growth rates of the fastest unstable growing mode versus wavelength of the perturbations for different level of the magnetic strength and various sizes of dust particles. Here, the electrical charge of dust particles is assumed to be $|Z|=+1$. Top plot of Figure \ref{fig:f1} shows the dispersion relation for weak magnetic field, but the bottom plot is for  magnetic field with stronger strength. We noticed as the strength of the magnetic field increases, the time-scale of the growing modes becomes independent of the variations of mass of the dust particles. In top plot of this figure, solid curves are corresponding to magnetic strength $B=10$ $\mu$G, but dashed lines are for $B=20$ $\mu$G. Each curve is labeled by the size of the dust particles.  When the size (i.e., mass) of the dust particles increases, the time-scale of the growing modes increases too. In other words, mass of the dust particles has a stabilizing effect on the growing modes.

In bottom plot of Figure \ref{fig:f1}, the dispersion relation is shown for stronger magnetic field strengths, i.e.  $B=60$, 80 and 100 $\mu$G. However, the plots are insensitive to the mass of small dust particles. We see as the magnetic field increases, the growth rate of the unstable modes at a particular wavelength of the perturbation increases. It means that the magnetic field has a destabilizing effect on the unstable modes.

\begin{figure}
\epsfig{figure=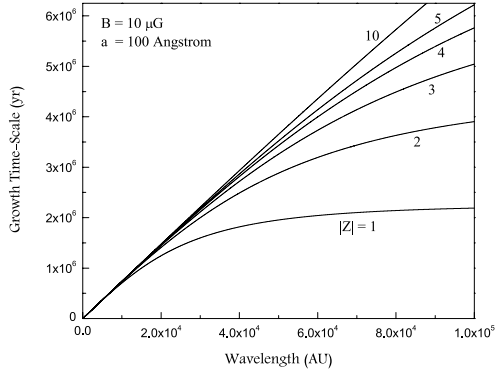,angle=0,width=8.0cm}
\caption{Growth time scales of the instability vs. wavelength of the perturbations for different electrical charge of the dust particles.}
\label{fig:f2}
\end{figure}

Figure \ref{fig:f2} shows the dispersion relation for the magnetic strength $B=10$ $\mu$G and dust particles with radius $100$ Angstrom, but changing the electrical charge of the grains. As the charge of the grain increases, the growing time-scale of the unstable modes increases.

We can understand the latter result as follows.  From the form of our
dispersion relation it is clear that the only variable which changes as
we vary the grain radius, $a$, is $f$ through the modified dust cyclotron
frequency, $\omega_{mcd}$.  Using relations (\ref{eq:nu}),
(\ref{eq:dn}), the definition of $\omega_{mcd}$ and recalling that the
mass of a dust grain is proportional to $a^3$ we can see that as $a$ increases
$\omega_{mcd}$ decreases.  Physically this means that larger grains are less
well coupled to the magnetic field, since the dust grains can only respond to
variations of the magnetic field with frequency less than $\omega_{mcd}$.
As Figure \ref{fig:f1} shows, the magnetic field destabilizes the flow.
Hence it is no surprise that as the dust component of the flow becomes less
coupled to the field the flow becomes more stable.  A similar argument
explains the results in Fig \ref{fig:f2}. According to the definition of the modified dust
cyclotron frequency, we have $\omega_{ mcd} \propto |Z|/\Theta$. Since  equation (\ref{eq:Theta}) shows that $\Theta \propto |Z|^{2}$, we can conclude that $\omega_{ mcd} \propto |Z|^{-1}$. Thus, for increasing values of $|Z|$ the dust cyclotron frequency $\omega_{mcd}$ decreases
which implies less magnetic coupling and greater growth times.  Interestingly,   numerical
simulations (e.g. Wiechen  2006) and laboratory experiments (e.g. Luo,
Angelo \& Merlino 2001) have also shown that the Kelvin-Helmholtz mode is
significantly stabilized with increasing mass of the dust.

\section{Discussion}

We studied the KH instability in a multifluid system analytically. Our results show that a higher mass of the dust particles has a stabilizing effect on the growing KH modes. But when the strength of the magnetic field increases, the time-scale of the growing modes gradually are becoming independent of dust size or mass. Also, the dynamics of unstable modes are independent from the charge polarity of the dust. Considering the wavelength of the perturbations and the time scale of the growing modes, we may conclude that some of structures which are seen in dusty outflows (e.g., Weinberger \& Armsdorfer 2004; Gueth, Bachiller \& Tafalla 2003; Shepherd 2001) are produced by the KH instability.

\section*{Acknowledgments}

We are grateful to  the referee, G. E. Ciolek,  for suggestions and comments to improve the paper.
The research of M. S. was funded under the Programme for Research in
Third Level Institutions (PRTLI) administered by the Irish Higher
Education Authority under the National Development Plan and with partial
support from the European Regional Development Fund.

{}

\section*{The coefficients of the dispersion equation}

The coefficients of equation (\ref{eq:dissal})  are

\begin{displaymath}
P_8 = M^4,
\end{displaymath}
\begin{displaymath}
P_7 = -8M^4,
\end{displaymath}
\begin{displaymath}
P_6 = -8M^6-4M^8+28M^4+4M^2f,
\end{displaymath}
\begin{displaymath}
P_5 = -20M^2f-56M^4+48M^6+8M^4f+8M^8,
\end{displaymath}
\begin{displaymath}
P_4 =-4M^6f+70M^4-20M^4f+4f^2+40M^2f
\end{displaymath}
\begin{displaymath}
+36M^8+8M^{10}+6M^{12}-120M^6,
\end{displaymath}
\begin{displaymath}
P_3 = -8\,{f}^{2}+32\,{M}^{10}+160\,{M}^{6}+8\,{M}^{12}-40\,{M}^{2}f
\end{displaymath}
\begin{displaymath}
+32\,{M}^{4}f-144\,{M}^{8}-56\,{M}^{4}+40\,{M}^{6}f-16\,{M}^{8}f,
\end{displaymath}
\begin{displaymath}
P_2 = 8\,{M}^{14}+8\,{M}^{2}{f}^{2}+40\,{M}^{8}f-120\,{M}^{6}-4\,{M}^{10}f
\end{displaymath}
\begin{displaymath}
-8\,{M}^{4}{f}^{2}+4\,{f}^{2}+36\,{M}^{12}+196\,{M}^{8}-4\,{M}^{16}+28\,{M}^{4}
\end{displaymath}
\begin{displaymath}
-40\,{M}^{4}f-144\,{M}^{10}-16\,{M}^{6}f+20\,{M}^{2}f,
\end{displaymath}
\begin{displaymath}
P_1 =  -120\,{M}^{8}+8\,{M}^{2}{f}^{2}+8\,f{M}^{12}-20\,{M}^{10}f
\end{displaymath}
\begin{displaymath}
-120\,{M}^{12}-4\,{M}^{2}f+48\,{M}^{6}+8\,{M}^{4}{f}^{2}+32\,{M}^{8}f
\end{displaymath}
\begin{displaymath}
+24\,{M}^{4}f-8\,{M}^{4}+48\,{M}^{14}+160\,{M}^{10}-8\,{M}^{16}
\end{displaymath}
\begin{displaymath}
-40\,{M}^{6}f,
\end{displaymath}
\begin{displaymath}
P_0 = -8\,{M}^{18}+{M}^{20}-20\,f{M}^{12}+4\,{M}^{14}f-8\,{M}^{6}
\end{displaymath}
\begin{displaymath}
+28\,{M}^{8}+4\,{M}^{4}{f}^{2}+28\,{M}^{16}+4\,{f}^{2}{M}^{8}-40\,{M}^{8}f
\end{displaymath}
\begin{displaymath}
+40\,{M}^{10}f-4\,{M}^{4}f+70\,{M}^{12}-8\,{f}^{2}{M}^{6}-56\,{M}^{14},
\end{displaymath}
\begin{displaymath}
-56\,{M}^{10}+{M}^{4}+20\,{M}^{6}f.
\end{displaymath}
where $f=(\omega_{mcd}/kU)^{-2}$.


\begin{thebibliography}{}

\bibitem[]{} Alton P. B., Davies J. I., Bianchi S., 1999, A\&A, 343, 51

\bibitem[]{} Birk G. T., Wiechen H., 2002, Physics Plasmas,  9, 964

\bibitem[]{} Birk G. T., Wiechen H., Lesch H., Kronberg P. P., 2000, A\&A, 353, 108


\bibitem[]{} Bodo G., Massaglia S., Rossi P., Rosner R., Malagoli A., Ferrari A., 1995, A\&A, 303, 281

\bibitem[]{} Chandrasekhar S., 1961, Hydrodynamics and Hydromagnetic Stability (Oxford: Oxford Univ. Press)


\bibitem[]{} Chhajlani R. K., Vyas M. K., 1991, Ap\&SS, 176, 69

\bibitem[]{} Chhajlani R. K., Vyas M. K., 1990, Ap\&SS, 173, 109

\bibitem[]{} Ciolek G. E., Mouschovias T. Ch., 1993, ApJ, 418, 774

\bibitem[]{} Downes T. P., Ray T. P., A\&A, 1998,  331, 1130

\bibitem[]{} Gueth F., Bachiller R., Tafalla M., A\&A, 2003, 401, L5

\bibitem[]{} Hardee P. E., Stone J. M., 1997, ApJ,  483, 121

\bibitem[]{} Hunter J. H., Whitaker R. W., 1989, ApJS, 71, 777

\bibitem[]{} Luo Q. Z., Angelo N. D., Merlino R. L., 2001, Physics Plasmas, 8, 31

\bibitem[]{} Markwick-Kemper F., Green J. D., Peeters E., 2005, 628, L119

\bibitem[]{} Mehta V., Bhatia P. K., 1988, Ap\&SS, 141, 151

\bibitem[]{} Pandey B. P., Vladimirov S. V., 2007, ApJ, 664, 942

\bibitem[]{} Rosen A., Hardee P. E., Clarke D. A., Johnson A., 1999, ApJ, 510, 136

\bibitem[]{} Roychoudhury S., Lovelace R. V. E., 1986, ApJ, 302, 188

\bibitem[]{} Shepherd D. S., ApJ, 2001, 546, 345

\bibitem[]{} Shadmehri M., Downes T. P., Astrophysics \& Space Science, 2007, in press

\bibitem[]{} Watson C., Churchwell E., Zweibel E. G., Crutcher R. M., 2007, ApJ, 657, 318

\bibitem[]{} Watson C., Zweibel E. G., Heitsch F., Churchwell E., 2004, ApJ, 608, 274

\bibitem[]{} Weinberger R., Armsdorfer B., 2004, A\&A, 416, L27

\bibitem[]{} Wiechen H., 2006, Physics Plasmas, 13, 062104





\end{thebibliography}
\end{document}